\documentclass[12pt]{article}
\usepackage[polish,english]{babel}
\usepackage[latin1]{inputenc}
\usepackage{times}
\usepackage[T1]{fontenc}
\usepackage{epsfig}

\begin{document}

\title{Note on scale invariance and self-similar evolution in (3+1)-dimensional signum-Gordon model }

\author{H. Arod\'z, $\;$ J. Karkowski \\ Institute of Physics,
Jagiellonian University\thanks{ Reymonta 4, 30-059 Cracow, Poland}
 \and Z. \'Swierczy\'nski \\ Institute of Computer Science and Computer Methods,\\ Pedagogical
University, Cracow\thanks{Podchor\c{a}\.{z}ych 2, 30-084 Cracow,
Poland }}

\date{$\;$}

\maketitle

\begin{abstract}
Several classes of self-similar, spherically symmetric solutions of
relativistic wave equation with a nonlinear term of the form
$\mbox{sign}(\varphi)$
 are presented. They are constructed from
cubic polynomials in the scale invariant variable $t/r$. One class
of solutions describes a process of wiping out the initial field,
another an accumulation  of field energy in a finite and growing
region of space.

\end{abstract}

\vspace*{2cm} \noindent PACS: 11.27.+d, 98.80.Cq, 11.10.Lm \\

\pagebreak

\section{ Introduction}

Symmetry  transformations for a given  equation act within the
space of its solutions. Especially interesting are fixed points of
such transformations, i.e., the solutions that remain invariant.
In the case of continuous symmetries they depend on a reduced
number of independent variables. This significantly simplifies the
task of finding the solutions, the task that so often is
formidably difficult in the case of nonlinear field equations.
Apart from the rudimentary examples of rotational and
translational symmetries, in certain models there is also a
one-parameter scaling symmetry. The corresponding invariant
solutions, called the self-similar ones, play an important role in
mathematical analysis of nonlinear evolution equations, as well as
in physics oriented investigations, see, e.g., the books \cite{1},
\cite{2} and a sample of recent research papers \cite{3}.

The present paper is a sequel to papers \cite{4}, where self-similar
solutions in the signum-Gordon model with a real scalar field in 1+1
dimensions were analyzed. Rich variety of such solutions was found.
Certain related models were investigated in \cite{5}. Derivation and
a general discussion of the signum-Gordon model can be found in
\cite{4}.  The model is very interesting, there is no doubt that its
investigations should be continued in several directions, such as
finding classical solutions of various kinds, applications, or
quantization.

As is well-known, field theories in 1+1 dimensions have rather
special properties. Therefore, it is not clear whether the amazing
picture of the space of self-similar solutions found in \cite{4} is
valid also in higher dimensions. This is our main motivation for the
present work. Another attraction is the possibility of obtaining
several exact analytic solutions to the signum-Gordon equation in
3+1 dimensions, and to have new insights into rather intriguing
dynamics of the scalar field in that model.

We have found several families of self-similar, spherically
symmetric analytic solutions. They are composed from cubic
polynomials in the  scale invariant variable $u=t/r$. Our results
show that there is a qualitative similarity between the 3+1 and 1+1
dimensional cases, but significant differences appear. First, the
lack of translational invariance in the radial variable $r$ results
in the absence of counterparts of certain classes of 1+1 dimensional
solutions.  Second difference, perhaps not so unexpected, is that
the pertinent calculations are a bit more complicated because we
have to look for zeros of cubic polynomials, while in the 1+1
dimensional case only the second order polynomials were present.
The exact solutions we have found provide examples of non-trivial
evolution of the scalar field $\varphi$. Especially interesting seem
to be solutions of the types $II$ and $III$, presented in Section 3,
which show  how the field settles exactly at the vacuum value
$\varphi=0$.

The plan of our paper is as follows. In  Section 2 we describe  the
method of constructing the self-similar solutions in the
signum-Gordon model. Explicit solutions  in the 3+1 dimensional case
are presented in Section 3. Section 4 is devoted to discussion of
our results.

\section{Scale invariant Ansatz and the method of constructing the solutions }

The signum-Gordon equation for the real scalar field $\varphi$ in
$D+1$ dimensional space-time reads \begin{equation}
\partial_{\mu}\partial^{\mu} \varphi(x) + \mbox{sign}(\varphi(x))
=0.
\end{equation}
The $\mbox{sign}$ function takes the values $\pm 1, 0$,
$\mbox{sign}(0) =0$. It is clear that Eq.\ (1) implies that the
second derivatives of $\varphi$ may not exist if $\varphi$ vanishes
at isolated points, but even at such points one-sided second order
derivatives do exist. The proper mathematical framework for
discussing equations of such kind is well-known: one should consider
the so called weak solutions, see, e.g.,  \cite{6}. All solutions
constructed below have been checked in this respect.

From a given solution $\varphi(x)$ of Eq.\ (1) one can obtain
one-parameter family of solutions of the form
\begin{equation}
\varphi_{\lambda}(x) = \lambda^2 \varphi(\frac{x}{\lambda}),
\end{equation}
where $\lambda >0$ is a constant. The solution is self-similar if
$\varphi_{\lambda}(x) = \varphi(x)$ for all $\lambda >0$. Taking
$\lambda=r$, where $r$ is the radius, and assuming the spherical
symmetry, we may then write
\begin{equation}
\varphi(x) = r^2\: G(\frac{t}{r}),
\end{equation}
where $t=x^0$ is the time. With this Ansatz, Eq.\ (1) gives
\begin{equation}
(u^2-1) G'' - (D+1)u G' + 2 D G - \mbox{sign}(G) =0,
\end{equation}
where $u=t/r$ and $G' = dG/du$. This equation has the particular
solutions
\[
G=0, \;\; G= \pm \frac{1}{2D}, \] which are trivial, but
nevertheless play important role below.

Let $g(u)$ be a solution of the auxiliary linear (!) equation
\begin{equation}
(u^2-1) g'' - (D+1)u g' + 2 D g  =0.
\end{equation}
Then $G_+(u) = g(u) + 1/(2D)$ is a solution of (4) on the interval
of $u$ defined by the condition $G_+(u) >0$, and $G_-(u) = g(u) -
1/(2D)$ on the interval in which $G_-(u) <0$. It turns out that
patching together a number of such partial solutions and, in some
cases, including also the trivial solution $G=0$, one can cover the
whole interval $0 \leq u < \infty$. In this way we obtain a
self-similar solution of Eq.\ (1) valid for all $r\in[0, \infty)$
and $t\in[0, \infty)$.  The values $G(0), G'(0)$ determine the
initial data for $\varphi(t,r)$: \begin{equation} \varphi(0,r) = r^2
G(0), \;\; (\partial_t\varphi)(0,r) = r G'(0).
\end{equation}
When patching the partial solutions we demand continuity of $G(u)$,
and also continuity of $G'(u)$, unless $u=1$. $G'(u)$ at $u=1$ does
not have to be continuous because $G''$ in Eq.\ (1) is multiplied by
the factor $u^2-1$ which vanishes at that point. This is not
surprising because $u=1$ corresponds to the light-cone $r=t$, the
characteristic hypersurface for equation (1).

Equation (5) has two linearly independent solutions:
\begin{equation}
g_1(u)= Du^2 +1, \;\;\; g_2(u) = \sum_{l=0}^{\infty} c_l u^{2 l +1},
\end{equation} where the coefficients $c_l$ are determined from the
recurrence relation
\begin{equation}
c_{l+1} = \frac{(2l-1) (2l+1-D)}{2 (l+1) (2l+3)} c_l.
\end{equation}
Note that in the case of odd space dimension $D$ the series for
$g_2$ is in fact a polynomial of order $D$, because $c_l=0$ for all
$l > (D-1)/2$. In particular, $g_2 = u$ for $D=1$, $g_2(u) = u^3 + 3
u$ when  $D=3$, and  $g_2= u^5 -10u^3 -15 u$ for $D=5$, apart from
overall multiplicative constants.

In the next Section we consider in detail the $D=3$ case.

\section{ Explicit self-similar solutions in the case $D=3$}

In the case of three space dimensions the general form of the
partial solutions reads
\begin{equation}
G= \pm \frac{1}{6} + \alpha (3 u^2 +1) + \beta(u^3 +3u),
\end{equation}
or equivalently,
\begin{equation}
G= \pm \frac{1}{6} + \gamma (u-1)^3 + \delta(u +1)^3,
\end{equation}
where  $\alpha, \beta, \gamma, \delta$ are constants. The sign in
front of 1/6 has to be equal to $\mbox{sign}(G)$.

Let us begin our exploration of the space of the self-similar
solutions by checking their asymptotic  behavior at $u \rightarrow
\infty$. It turns out that $G$ has to have a constant sign if values
of $u$ are large enough -- we prove in the Appendix  the following
lemma. \\
\textbf{Lemma.} Suppose that $u_1 >1$ is a real zero of the
polynomial (9). Then, that polynomial does not have any real zeros
larger than
$u_1$. \\
It follows that in the region $u \geq u_1$ the solution has the
fixed form (9) with certain fixed coefficients $\alpha, \beta$.
The just excluded option was that of having an infinite sequence
of polynomials with alternating signs. Now, let us recall that the
region $u\rightarrow \infty$ corresponds to  $ r \rightarrow 0$
($t>0$). Therefore, the polynomial $u^3 + 3u$ would give a
singular field $\varphi = r^2 G \sim 1/r$ with a divergent, non
integrable at $r=0$ energy density. For this reason, we put $\beta
=0$. Note also that by taking into account the symmetry $\varphi
\rightarrow - \varphi$, we may consider just two cases of the
asymptotic form of $G$ at large values of $u$:  $G(u)
>0$, or $G(u)=0$.  The latter case is considered in the second and
third subsections (solutions of the types II and III).  The former
case is named type I.

\subsection{Solutions of type $I$}

In this case, for arbitrary large values of $u$
\[
G(u) = \frac{1}{6} + \alpha (3 u^2 +1) >0,
\]
as pointed out in the preceding paragraph. This implies that
$\alpha \geq 0$, and then such $G(u)$ does not have any zeroes
down to $u=1$, where it can be glued with another polynomial of
the form (9) or (10). We conclude that all type I solutions on the
interval $u \in [1, \infty)$ have the form
\begin{equation}
G_{\infty}(u) = \frac{1}{6} + \alpha_{\infty} (3 u^2 +1)
\end{equation}
with an arbitrary constant $\alpha_{\infty} \geq 0$.

At the point $u=1$ the solution $G_{\infty}$ can be glued with
another polynomial (9) or (10), taken with the $+1/6$ because
$G_{\infty}(1)= 1/6+ 4 \alpha_{\infty} >0$. Let us denote such
polynomial by $G_+$ and its constants by $\gamma_+, \delta_+$
(here we prefer the form (10)). The gluing condition at $u=1$,
$G_{\infty}(1) = G_+(1)$ gives $\delta_+= \alpha_{\infty}/2$.
Hence,
\begin{equation}
G_+(u)=  \frac{1}{6} + \gamma_+ (u-1)^3 + \frac{1}{2}
\alpha_{\infty} (u +1)^3.
\end{equation}
Now we have to determine the lower end of the interval on which
$G_+$ is the solution (the upper end is $u=1$). The first
possibility, denoted as $Ia$, is that $G_+ >0$ for all $u\in[0,1)$.
This occurs if $\gamma_+ < 1/6 + \alpha_{\infty}/2$. In this case
$G_+$ and $G_{\infty}$ cover the whole interval $[0, \infty)$, and
these functions together give the complete solution. It has the
shape sketched in Fig.\ 1 with the dashed line.

\begin{center}
\begin{figure}[tph!]
\hspace*{1cm}
\includegraphics[height=9cm, width=9cm]{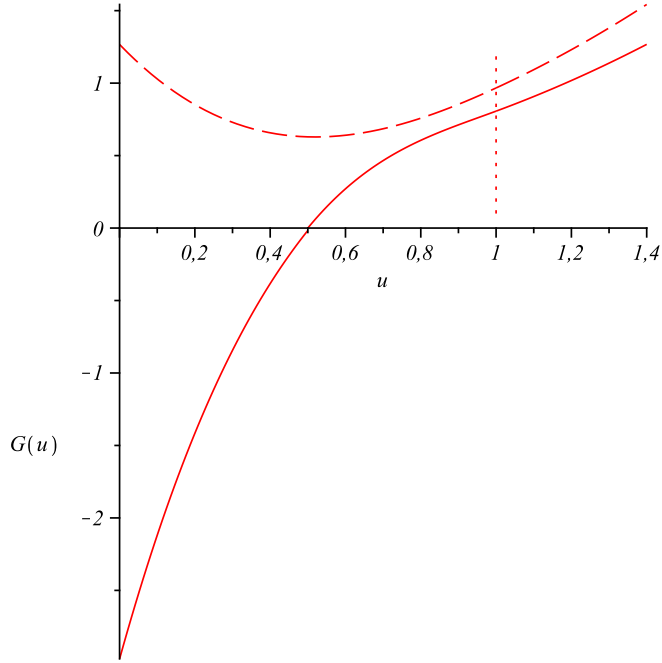}
\caption{Solutions of the type $Ia$ (the dashed line) and $Ib$
(the solid line)}
\end{figure}
\end{center}

The other possibility, denoted as $Ib$, is that there exists $u_0
\in (0,1)$ at which $G_+(u_0)=0$. Simple calculation gives in this
case
\begin{equation}
\gamma_+ =  \frac{1 + 3 \alpha_{\infty} (1+u_0)^3}{6 (1-u_0)^3}.
\end{equation}
Moreover, it turns out that $u_0$ has to be the first order zero of
$G_+$. Therefore, at the point $u_0$ we may glue $G_+$ with a
negative polynomial
\begin{equation}
G_- = -\frac{1}{6} + \gamma_- (u-1)^3 +  \delta_- (u +1)^3  \leq 0,
\end{equation}
and not with the trivial solution $G=0$. The matching conditions
\[
G_-(u_0)=0, \;\;\; G'_-(u_0) = G'_+(u_0)
\]
give
\begin{equation}
\gamma_- =  \frac{u_0 + 3 \alpha_{\infty} (1+u_0)^3}{6 (1-u_0)^3},
\;\; \delta_-= \frac{\alpha_{\infty}}{2}  + \frac{1}{6 (1+u_0)^2}.
\end{equation}

It turns out that $G_-(u)$ found above does not have any zeros in
the interval $[0, u_0)$. Therefore, the three functions:
$G_{\infty}$ for $u \geq 1$, $\;G_+$ for $u_0 \leq u \leq 1$, and
$G_-$ for $0 \leq u \leq u_0$, form the complete solution of Eq.\
(4). It is sketched in Fig.\ 1 with the solid line.

The values of $G(0), G'(0)$, which specify the initial values of the
scalar field through formula (6), read as follows. \\
Type $Ia$: \[ G_+(0) = \frac{1}{6} - \gamma_+ +
\frac{\alpha_{\infty}}{2}, \; \; G'_-(0) = 3 \gamma_+ + \frac{3
\alpha_{\infty}}{2}. \] Type  $Ib$:
\[
G_-(0) = -\frac{1}{6} - \gamma_- + \delta_-, \;\; G'_-(0)=
3(\gamma_- + \delta_-). \] Varying $\alpha_{\infty}$ in the interval
$[0, \infty)$ and $u_0$ in the interval $[0,1)$ we obtain certain
sets in the $(G(0), G'(0))$  plane. They are shown in Fig.\ 2. We
have denoted them  $Ia, Ib $, identically as the corresponding
solutions.

The regions $-Ia, -Ib $ are obtained by the reflection in the
origin, $ (G(0), G'(0)) \rightarrow  (-G(0), -G'(0))$, related to
the symmetry $\varphi \rightarrow - \varphi$. The half-infinite
curve that separates the region $Ib$ from $IIb$ is obtained for
$\alpha_{\infty} =0$. It has the following parametric form with
$u_0 \in [0, 1)$ as the parameter:
\[ G(0)=  \frac{1}{6 (1+u_0)^2} -
\frac{u_0}{6 (1-u_0)^3} - \frac{1}{6}, \;\;\; G'(0)= \frac{1}{2
(1+u_0)^2} + \frac{u_0}{2 (1-u_0)^3}. \] The curve starts at the
point $(0, 1/2)$ that corresponds to $u_0=0$, and it approaches
its asymptote, that is the straight half-line
\[G(0)(\sigma) = - \frac{1}{8} - \sigma, \;\;\; G'(0)(\sigma) = \frac{1}{8} + 3 \sigma,
\;\;\;  \sigma \in [0, \infty), \] when $u_0 \rightarrow 1$.

\begin{center}
\begin{figure}[tph!]
\hspace*{1cm}
\includegraphics[height=11cm, width=11cm]{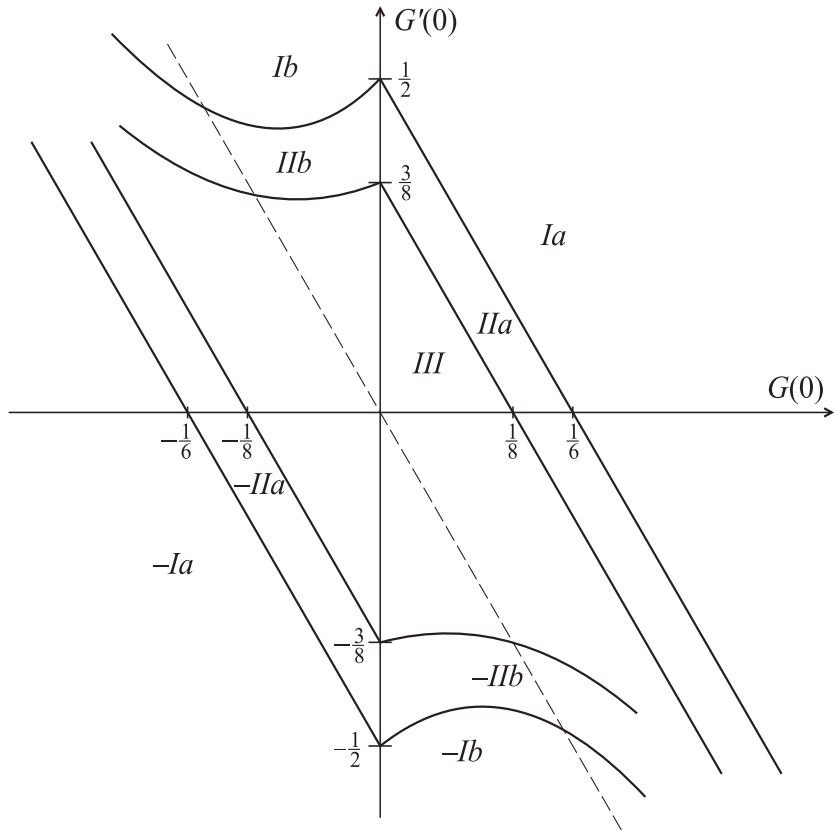}
\caption{The map of the self-similar solutions. The regions $Ia$,
$IIa$  are separated from $Ib$, $IIb$, respectively, by the
$G'(0)$ axis. The infinite region $III$ extends over all area
encompassed by the stripes $IIa, IIb, -IIa, -IIb$, except for the
origin $(0,0)$ which corresponds to the trivial solution $G=0$.
The stripe $IIb$ asymptotically approaches the stripe $-IIa$,
similarly $IIa$ approaches $-IIb$.  The meaning of the  dashed
straight line is explained in Section 4}
\end{figure}
\end{center}

\subsection{Solutions of type $II$}

Solutions of the type $II$ are obtained by taking the trivial
solution $G(u)=0$ for $u\geq u_1 >1$, and gluing it at the point
$u_1$ with the polynomial (10) taken with the $+$ sign (the other
sign can be obtained by the symmetry transformation $\varphi
\rightarrow - \varphi$). Because $u_1 \neq 1$, the matching
condition includes the first derivative, \[ G(u_1)=0, \;\; G'(u_1)
=0.\] Solving these conditions we obtain the following polynomial
\begin{equation}
G_d(u) = \frac{(u_1-u)^2 (2 u_1 u + u_1^2 -3)}{6 (u_1^2-1)^2}.
\end{equation}
It has strictly positive values in the whole interval $u \in[1,
u_1)$. At $u=1$ that polynomial is glued with another positive
polynomial $G_+(u)$ of the form (9). The matching condition $G_d(1)
= G_+(1)$ gives
\begin{equation}
G_+(u) = \frac{1}{6} - \alpha_+ (u-1)^3 - \frac{u (u^2
+3)}{6(1+u_1)^2},
\end{equation}
where $\alpha_+$ is an arbitrary constant. Similarly as for the type
$I$ solutions, there are two possibilities. The first, denoted as
$IIa$, with $G_+(u)$ strictly positive on the whole interval
$(0,1]$, occurs when $\alpha_+ \geq -1/6$. In this case we already
have the complete solution: $G=0$ for $u \geq u_1$, $G_d$ for $u \in
[u_1, 1]$, and $G_+$ given by formula (17) for $u\in [0,1]$. It is
sketched in Fig.\ 3 with the dashed line. The values of $G_+, G_+'$
at $u=0$, i.e.,
\[
G_+(0) = \frac{1}{6}+ \alpha_+, \;\; G_+'(0) = - 3 \alpha_+ -
\frac{1}{2(1+u_1)^2}, \] where $\alpha_+ \in [-1/6, \infty), \; u_1
\in (1, \infty)$, fill  a semi-infinite straight-linear stripe in
the $(G(0), G'(0))$ plane. It is denoted as $IIa$ in Fig.\ 2.

The second possibility is that $G_+(u)$ has a zero at some $u_0\in
(0,1)$. If this is the case, then
\[ \alpha_+ = \frac{u_0(u_0^2+3)}{6 (1-u_0)^3 (1+u_1)^2} -
\frac{1}{6 (1-u_0)^3}. \] Formula (17) implies that the zero can
only be of the first order. The matching conditions at $u_0$
determine the two coefficients in the negative polynomial
\begin{equation}
G_-(u) = - \frac{1}{6} + \gamma_- (u-1)^3 + \delta_- (1+u)^3
\end{equation}
which is glued at that point with $G_+(u)$. Simple calculations give
\[
\gamma_- = \frac{u_0}{6 (1-u_0)^3} - \frac{(1+u_0)^3}{12 (1 +u_1)^2
(1-u_0)^3}, \;\; \delta_-= \frac{1}{6 (1+u_0)^2} - \frac{1}{12
(1+u_1)^2}.
\]

It turns out that $G_-(u)$ is strictly negative in the whole
interval $u\in [0, u_0)$ for all choices of $u_0 \in (0, 1), \;
u_1 \in (1, \infty)$. Thus, we have obtained again the complete
solution:  $G=0$ for $u \geq u_1$, $G_d$ for $u \in[1,u_1]$, $G_+$
for $u \in [u_0,1]$, and $G_-$ for $u \in [0,u_1]$. We denote it
as $IIb$.  It is sketched in Fig.\ 3 with the solid line.

\begin{center}
\begin{figure}[tph!]
\hspace*{1cm}
\includegraphics[height=9cm, width=9cm]{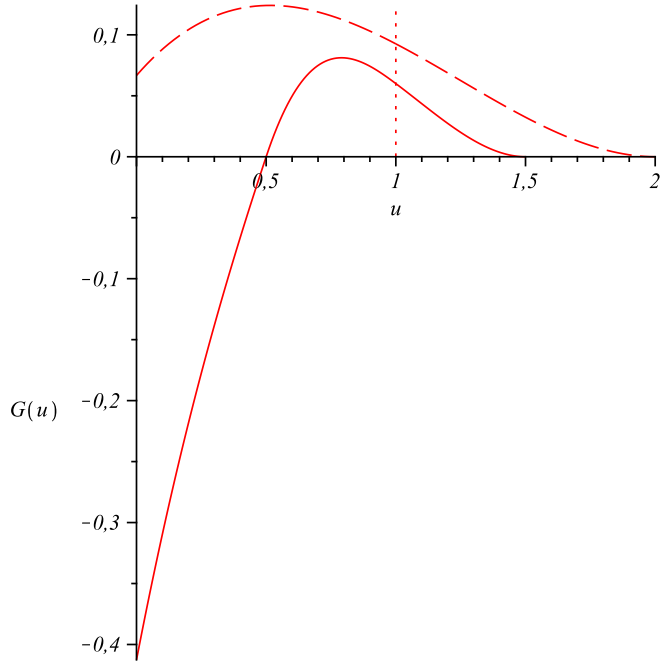}
\caption{Solutions of the type $IIa$ (the dashed line) and $IIb$
(the solid line) }
\end{figure}
\end{center}

The points $(G_-(0), G'_-(0))$ with $u_0\in(0,1), u_1 >1$ fill in
the $(G(0), G'(0))$ plane the region denoted as $IIb$, see Fig.\ 2.
It borders  the region $IIa$ along the segment $(3/8, 1/2)$ of the
$G'(0)$ axis. The region $IIb$  is bounded from below by the curve
obtained by putting $u_1=1$ in formulas for $G_-(0), G'_-(0)$ for
the solution (18). The curve is obtained in the parametric form,
namely
\[ G_-(0) =  - \frac{u_0(2+u_0)}{6 (1+u_0)^2} - \frac{u_0(1+u_0)}{24
(1-u_0)^2}, \;  G_-'(0) = \frac{1}{2(1+u_0)^2} +
\frac{3u_0-1}{8(1-u_0)^2}, \] where $u_0$ varies in the interval
$(0, 1)$. The asymptote of this curve, approached when $u_0
\rightarrow 1-$, coincides with the upper boundary of the region
$-IIa$ in Fig.\ 2 (the semi-infinite straight line that starts from
the point $(0, -3/8)$ on the $G'(0)$ axis).

\subsection{Solutions of type $III$}

Thus far we have not found solutions for which the points $(G(0),
G'(0))$ would lie in the central part of the map presented in Fig.\
2. In fact, we have not exhausted yet all possibilities for having
the solutions of the type $II$ because we have assumed that $u_1 >
1$.  Simple calculations show that the partial solution of the form
$G_d$, formula (16),  does not exist if $u_1 <1$. Thus, we are left
with the last possible choice, namely $u_1 =1$. In this case
$G(u)=0$ for $u \geq 1$.  The matching condition with the trivial
solution $G=0$ at $u=1$ does not involve the first derivative. For
his reason, we regard such solutions as essentially different from
the ones already constructed, and we classify them as the type
$III$.

As for the general form of such solutions we are guided by the
results obtained for the signum-Gordon model in 1+1 dimensions,
\cite{4}. Thus, in the present case we expect an infinite sequence
$\{G_1, G_2, G_3, \ldots \}$ of polynomials of the form (10) with
alternating signs, glued together at the points $u_0, u_1, u_2,
\ldots $ lying in the interval $(0,1)$ and converging at $u=1$.
These points are the first order zeros of the pertinent
polynomials, and the matching conditions include the first
derivatives. Unfortunately, we have not been able to fully
construct these solutions in exact, analytic form because the
cubic polynomials pose significant problems in explicit
calculations. One can relatively easily construct several first
polynomials with the desired properties, but the infinite sequence
of cubic polynomials is a different matter.

Numerical calculations support our expectations. Example of such a
numerical solution is shown in Fig.\ 4. It has been calculated up to
the point $u=0.9996$, determined by numerical accuracy of the
computation.  Of course, the numerical results for this kind of a
problem, where we need to check an infinite sequence of polynomials,
are not decisive, even if they are quite suggestive. For this
reason, the given above description of the solutions of the type
$III$ has, strictly speaking,  the status of a conjecture, as
opposed to the status of the solutions of the types $I$ and $II$.

\begin{center}
\begin{figure}[tph!]
\hspace*{1cm}
\includegraphics[height=7cm, width=11cm]{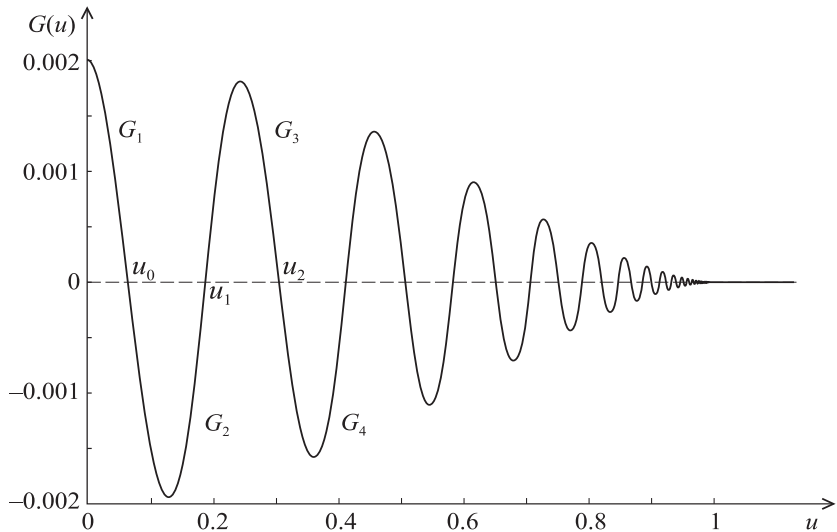}
\caption{Example of numerical solution of the type $III$ }
\end{figure}
\end{center}

\section{Discussion}

\noindent 1. We have found the three families of self-similar
solutions of the signum-Gordon equation in the $D=3$ case. Comparing
with the analogous results in 1+1 dimensions \cite{4}, we see
certain similarity. In particular, in both cases there exist
solutions of the type $III$ (i.e., solutions vanishing for all
$u\geq 1$) and of the type $II$ (vanishing for all $u \geq u_1 >1$).
This seems to be a universal feature of the signum-Gordon equation.
Solutions of the type $I$ were not found in \cite{4}, but this may
be a consequence of certain ad hoc restrictions on the form of
solutions adopted there.   On the other hand, in the $D=3$ case we
have not found spherically symmetric solutions that would correspond
to the presented in \cite{4} solutions with a negative velocity $v$.
This can be related to the lack of translational
invariance in the radial coordinate $r$. \\

\noindent 2.  The solutions presented above describe quite
interesting processes, especially the type $II$ solutions. The
region $u \geq u_1 >1$ in which $\varphi =0$ corresponds to $r \leq
t/u_1$. Thus, the field retreats completely from the region
surrounding the origin. The radius of the spherical region with the
vanishing field increases linearly with time. The corresponding
radial velocity is equal to $1/u_1$, and it is smaller than 1. The
presence of such subluminal velocities shows that actually the model
should not be regarded as massless, in spite of the presence of the
scale invariance. In truly massless models, such as the free
massless scalar field discussed in the next point, wave fronts
always move with the velocity $\pm 1$.

The type $I$ solutions give $\varphi(r,t) = (1/6  + \alpha_{\infty})
r^2 + 3 \alpha_{\infty} t^2$ in the region $r \leq t$. Thus, in this
case we have an accumulation of the field energy around the origin.
\\

\noindent 3.  It is interesting to compare the solutions described
in Section 3 above with self-similar radial solutions of the free
wave equation which is obtained by dropping the
$\mbox{sign}{\varphi}$ term from Eq.\ (1). It is clear that the
general solution of the free wave equation is given by  linear
combination of the functions $g_1(u), g_2(u)$  introduced in
Section 2.    It turns out that solutions of the type II do not
exist. Solutions of the type III have the form $G(u) = \alpha
(1-u)^3$ for $u\leq 1$, $G(u)=0$ for $u\geq 1$, where $\alpha $ is
an arbitrary constant. On the plane $(G(0), G'(0))$ these
solutions are represented by the points lying on the straight line
$G'(0) = -3 G(0)$, shown as the dashed line in Fig.\ 2. All other
points in that plane  would correspond in the case of the free
wave equation to solutions of the type I, which have the following
form: $ G(u)= 2 \beta (3 u^2 +1) $ for $u\geq 1$ and $G(u) = \beta
(1+u)^3 + \alpha (1-u)^3$ for $u \leq 1$, where $\beta \neq 0$.
Thus, the rather nontrivial shape of the region occupied by the
solutions of the type III in Fig.\ 2, as well as the presence of
the regions $IIa, IIb$, reflect the presence of the
$\mbox{sign}(\varphi)$ term.

\section{Appendix. The proof of Lemma}

\noindent 1.  If $u_1$ is the second order zero, the polynomial has
the form (16) modulo the overall sign. The factor $2 u_1 u + u_1^2
-3 $ is strictly positive for all $u >u_1$ because $u_1 >1$.
Therefore $G_d$ does not vanish for any $u > u_1$. \\

\noindent 2. In the case $u_1$ is the first order zero, we have
$G(u_1)=0,\; G'(u_1)\neq 0$. Because of the symmetry $\varphi
\rightarrow - \varphi$ we may consider only the case $G'(u_1) >0$.
Then, for $u \geq u_1$ we have the polynomial
\[
G_+(u) = \frac{1}{6} - \frac{1 + 3 u^2}{6(1+3 u_1^2)} + \beta_+
(u_1^3 + 3 u_1) \left[ \frac{u^3 + 3u}{u_1^3 + 3u_1} - \frac{1
+3u^2}{1+3 u_1^2}\right],  \] where
\[
\beta_+ > \frac{u_1}{3 (u_1^2-1)^2} \] in order to ensure that
$G'_+(u_1) >0$.  Direct calculation shows that $G''_+(u) >0$ for
all $u >u_1$. Therefore $G'_+(u) >0$ for all $u>u_1$, and in
consequence the values of $G_+(u)$ do not return to 0 in the
region $u>u_1$.


\begin{thebibliography}{99}
\bibitem{1} L. Debnath, \emph{Nonlinear Partial Differential Equations}. Birkh$\ddot{\mbox{a}}$user, Boston-Basel-Berlin, 2005. Section
8.11.
\bibitem{2} G. I. Barenblatt, \emph{Scaling, Self-similarity,
and Intermediate Asymptotics}. Cambridge University Press,
Cambridge-New York-Melbourne, 1996.
\bibitem{3} P. Bizo\'n, Commun. Math. Phys. \textbf{215}, 45 (2000); P.
Lauren\c{c}ot, Physica D \textbf{222}, 80 (2006); P. Bizo\'n, D.
Maison and A. Wasserman, Nonlinearity \textbf{20}, 2061 (2007);  M.
Herrmann, B. Niethammer and J. J. L. Vela\'{z}quez, J. Diff. Eq.
\textbf{247}, 2282 (2009); P. Biler, G. Karch and R. Monneau,
Commun. Math. Phys. \textbf{294}, 145 (2010); C. B. Muratov, P. V.
Gordon and S. Y. Shvartsman, Phys. Rev. E \textbf{84}, 041916
(2011).
\bibitem{4}  H. Arod\'z, P. Klimas and T. Tyranowski, Phys. Rev. E
\textbf{73}, 046609 (2006); H. Arod\'z, P. Klimas and T.
Tyranowski, Acta Phys. Pol. \textbf{B38}, 3099 (2007).
\bibitem{5} H. Arod\'z, P. Klimas and T. Tyranowski, Acta Phys. Pol. \textbf{B38}, 2537 (2007).
\bibitem{6} L. C. Evans, \emph{Partial Differential Equations}. American
Math. Society, Providence, RI,  1998.

\end{thebibliography}
\end{document}